\documentstyle[tighten,pra,aps,epsfig]{revtex}
\begin{document}
\title{Discrete Phase Measurements and the Bell inequality}
\author{W. J. Munro\cite{WJM}}
\address{Centre for Laser Science,Department of Physics, University of
Queensland,\\ QLD 4072, Brisbane, Australia}
\date{\today}
\maketitle
\begin{abstract}
We present a formulation of the Bell inequalities using simple 
correlated photon number states and phase measurements. Such 
tests generally require binning of the information, and this effect is 
closely examined. Our proposal opens up the opportunity for a new novel 
test of quantum mechanics versus local realism. Some insight in entanglement 
in such systems may be achieved.
\end{abstract}
\pacs{}

\section{Introduction}

The Bell inequalities\cite{Bell65,CHSH69,Clauser and Horne 74} and their 
tests of quantum mechanics have generally been seen only as a
fundamental test of quantum mechanics\cite{EPR35} until recently. 
Now however their relevance has become much more 
important with the proposal for quantum computing and quantum 
encryption. Fundamental to these schemes is quantum entanglement which is 
at the core of quantum mechanics. The Bell inequalities provide a 
mechanism now by which strongly\cite{strong} entangled two mode systems can be 
distinguished from all classical systems. Current tests of the Bell 
inequalities have suffered from a number of loopholes
\cite{CHSH69,Kwiat et al 94,Freyberger et ak 95,Fry et al 96}, including 
the fair sampling assumption and low detection 
efficiencies. To date, no definitive experiment test has been 
achieved although specific experiments have been performed which close 
one or more of the loopholes. This has lead to proposals from a number of authors including ourselves
\cite{Reid 98,Gilchrist Deuar and Reid 98,Munro 99,Munro and  Milburn 98,Yurke and Stoler 97} 
to consider high efficiency detection. Work by Smithery {\it et. al.}\cite{Smithey 1992} 
has indicated how high efficiency photon detectors can be 
used to detect large photon numbers but with an uncertainty in the 
actual photon number.  Reid {\it et. al.}\cite{Reid 98} indicated how high efficiency photon 
detectors, with large particle numbers, can test macroscopic quantum mechanics.

In a recent paper by Gilchrist {\it et. al.}\cite{Gilchrist Deuar and Reid 98} they showed how 
the predictions of quantum mechanics are in disagreement 
with those of local hidden variable theories for a situation 
involving continuous quadrature phase amplitude (ÒpositionÓ and ÒmomentumÓ)
measurements. More explicitly they showed that the quantum predictions
for the probability of obtaining results $x$ and $p$ for
position and momentum (or various linear combinations) cannot be predicted by any local hidden
variable theory. The test could be achieved by binning the continuous position and momentum 
information into two and using the binary results in the strong Clauser Horne Bell 
inequality test\cite{Clauser and Horne 74}. There predicted 
violation was small (less than 2\%). 
Munro \cite{Munro 99} considered various different strong Bell inequalities 
and indicated how larger violations may be achieved.

Also published at a similar time by us \cite{Munro and Milburn 98} we proposed how
the novel use of phase measurements on a simple correlated 
photon number triplet  could be used to test the GHZ correlations via 
the Mermin $F$ inequality\cite{Mermin90}. More explicitly we showed how simple correlated photon
number triplets which ideally could be produced in nondegenerate parametric 
oscillation, where we have signal, idler and pump modes, in 
conjunction with discrete binary phase measurements could be used to 
provide a definitive test of quantum mechanics versus all local realism.
In fact a maximal violation of the mermin $F$ 
inequality was indicated. However as we mentioned in this paper 
discrete phase measurements in optical systems have yet to be experimentally realized 
in the ultra high detector efficiency limit. As an approximation to 
this binary phase measurement a homodyne quadrature phase amplitude measurements 
was considered and also found to violate the inequality. The potential 
violation is significantly reduced however because information in the 
quadrature phase amplitude measurements must be discarded to achieve 
the binary result. 

In this paper we will be restricting our attention to systems 
involving correlated photon number states. To obtain the maximum entanglement 
information we are proposing the use of canonical phase measurements. 
Initially we consider a discrete binary phase measurement but then 
generalise to phase measurements involving larger (greater than two) 
numbers of phase results. To achieve a Bell inequality test in such 
cases requires binning of the results from the phase measurements.
A critical part of this paper is to determine the effect of this 
binning process. Until recently only two limits have been 
considered, binary discrete phase measurements and continuous 
quadrature phase amplitude measurements. We analyse a number of cases 
between these extreme limits. While our phase measurements may not be 
easily experimentally achievable valuable insight into the binning of 
data for the Bell inequalities is obtained. 
While we will not explicitly consider correlated spin systems, the results 
indicated by the binning of phases could be applied to these spin 
systems.  For instance with binning a correlated $3/2$ system could be used to test the Bell 
inequality. Historically only correlated $1/2$ systems 
have been used. This opening the possibility for new novel tests of quantum mechanics 
with higher spin particles.

\
\section{Entanglement}

It is important to begin by explaining the reason for considering 
phase measurements especially when we are restricting our 
attention to  correlated photon number state systems. 
As has long been known quantum entangled states shows correlations that cannot be explained
in terms of the correlations between local classical properties of 
the subsystems. In this paper we are describing a pure entangled state of 
two modes in which the correlations are in photon number, that is 
the nature of correlation can be succinctly stated by
saying that there are equal number of photons in each mode. Now as there
are many different ways to realize this, the total state is the sum
over amplitudes for all possible ways in which this correlation can be realized. 
This kind of sum over amplitudes for correlations is characteristic of an entangled
state.

How do we best see the quantum nature of the correlation? Obviously it is not 
enough to measure photon number as this would not distinguish a mixed state,
with equal photon numbers in each mode, from the equivalent entangled 
pure state. In some sense we need to measure an observable which carries as little information as
possible about photon number in order to see the interference 
between all the possible ways in which the correlation in photon number can be realized. 
We conjecture that the best choice is the observable canonically conjugate to photon number; 
the phase. Before we consider the quantum mechanical phase (Section \ref{phasesection}) 
we will in the next section specify more precisely the nature of the 
entangled states we are considering and realisable systems that 
generate them.

\section{Entangled Photon Pair States}

Tests of quantum mechanics generally require an entanglement between 
particles. A test of  the Bell inequality requires an entanglement 
between two subsystems. As mentioned previously we are 
examining a two mode state where there are an equal number of photons in each mode. 
This correlated photon number pair state is given by 
\begin{eqnarray}\label{correlatedpair}
| \Psi \rangle = \sum_{n=0}^{\infty} c_{n} | n \rangle | n \rangle 
\end{eqnarray}
where at present we leave $c_{n}$ arbitrary but specify that due to 
normalisation
\begin{eqnarray}
\sum_{n=0}^{\infty} \left|c_{n}\right|^{2}=1.
\end{eqnarray}
For convenience in the calculations that follow, we will place the 
condition on $c_{n}$ that it must be real. Eqn (\ref{correlatedpair}) 
actually describes a number of idealised 
but real systems. The most well known and extensively examined case is 
the nondegenerate parametric amplifier specified 
by an ideal  Hamiltonian of the form\cite{Reid 89}  
\begin{eqnarray}
H=-\hbar \chi \epsilon \left( a b+a^\dagger b^\dagger\right).
\end{eqnarray}
where $\epsilon$ is field amplitude of a nondepleting classical pump 
and  $\chi $ is proportional to the susceptibility of the medium.
$a,b$ are the boson operators for the two spatially separated orthogonal signal 
and idler modes systems. After a time $\tau$, the state of the system is given 
by (\ref{correlatedpair}) with $c_{n}$ specified by
\begin{eqnarray}\label{parampcn}
c_{n}={{\tanh^n \left[\chi \epsilon \tau\right]}\over{\cosh \left[\chi \epsilon \tau\right] }}
\end{eqnarray}
For small $\chi \epsilon \tau$ in Eqn (\ref{parampcn}) a state of the 
form
\begin{eqnarray}\label{twostatesystem}
| \Psi \rangle = c_{0}| 0 \rangle | 0 \rangle+ c_{1}| 1 \rangle | 1 \rangle  
\end{eqnarray}
can be generated as a  reasonably good approximation.

Another source of highly correlated photon number states exists in the steady state by 
nondegenerate parametric oscillation \cite{Reid and Krippner 93}
as modeled by the following Hamiltonian,
\begin{eqnarray}\label{hamcircle}
H=-i\hbar \epsilon \left( a b-a^\dagger b^\dagger\right)+a b 
\Gamma^{\dagger}+a^\dagger b^\dagger \Gamma.
\end{eqnarray}
Here we have neglected the effect of linear single photon loss. 
We assume that the coupled signal-idler loss dominates over linear
single-photon loss. In Eqn. (\ref{hamcircle}) $\epsilon$ represents a 
coherent driving source which generates signal-idler pairs, while 
$\Gamma$ represents the reservoir systems which gives rise to the coupled signal-idler loss.
$a,b$ again are the boson operators for the orthogonal signal and idler modes.
In the limit of very large parametric 
nonlinearity and high Q cavities, a state of the form\cite{Reid and Krippner 93} 
\begin{equation}
| {\mbox{circle}}\rangle = {\mathcal{N}} \int_{0}^{2\pi}    
|r e^{i\varsigma}\rangle_{a} |r e^{-i\varsigma}\rangle_{b} d\varsigma 
\end{equation}
can be generated. Here ${\mathcal{N}}$ is the normalisation coefficient 
while  $|{\ldots}\rangle_{a}$ and $|{\ldots}\rangle_{b}$ are coherent states 
of amplitude $r$ in the spatially separated modes $\hat a$ and $\hat b$. 
This state can be written in the form of 
eqn (\ref{correlatedpair}) with the $c_{n}$'s now specified by
\begin{eqnarray}\label{circlecn}
c_{n}={{r^{2 n}}\over{n! I_{0}\left(2 r^{2} \right)}} 
\end{eqnarray}
where $I_{0}$ is the zeroth order modified Bessel function.  

Given that we now have exactly specified the nature of the our 
correlation we return to consider the  observable canonically conjugate to photon number; 
the phase.

\section{Phase states}\label{phasesection}

Above we have proposed that that phase states may be the best way to 
observe entanglement in photon number. There have been a  number of 
phase states proposed over recent years\cite{Helstrom 76,Holevo 82,Braunstein 
96,Pegg and Barnett 88,Pegg and Barnett 89}. 
A canonical phase state\cite{Helstrom 76,Holevo 82,Braunstein 96} 
$|\theta \rangle$ can be generally be defined as
\begin{eqnarray}
| \theta \rangle =  {1 \over \sqrt{2 \pi}} 
\sum_{n=0}^{\infty} \exp \left[i n \theta \right] | n \rangle 
\end{eqnarray}
where $ | n \rangle$ represent the fock states and $\theta$ specifies 
the phase. These phase states are unnormalized as $\langle \theta| \theta\rangle >1$. In fact they are 
unnormalisable. A normalizable phase state was proposed by  
Pegg and Barnett\cite{Pegg and Barnett 88,Pegg and Barnett 89} who considered the 
phase state definition
\begin{eqnarray}\label{1}
| \theta \rangle = \lim_{s\rightarrow \infty} {1 \over \sqrt{s+1}} 
\sum_{n=0}^{s} \exp \left[i n \theta \right] | n \rangle. 
\end{eqnarray}
Here $| n \rangle$ are the $s+1$ number states that span the $(s+1)$ 
dimensional state space. The limit in Eqn~(\ref{1}) is absolutely necessary in 
order to normalize the states. The parameter $\theta$ can take on any real value, 
although distinct states will occur in a $2 \pi$ range. 

Pegg and Barnett\cite{Pegg and Barnett 88,Pegg and Barnett 89}  showed that a set of $s+1$ orthonormal phase 
states, with values of $\theta$ differing by $2 \pi /(s+1)$, can be 
generated from
\begin{eqnarray}\label{2}
| \theta_{\mu} \rangle = \exp \left[i \hat N \mu 2 \pi /(s+1) \right]
| \theta_{0} \rangle,\;\; \mu=0,\ldots,s
\end{eqnarray}
where $|\theta_{0} \rangle$ is the reference (or zero) phase state. Here 
$\hat N$ is the number operator and $\mu$ the particular
discrete phase we are interested in.  $\mu$ generally varies in integer steps from $0$ to $s$. 
Our values  for $\theta_{\mu}$ are given by
\begin{eqnarray}
\theta_{\mu}=\theta_{0}+ {{ 2 \mu \pi}\over{(s+1) }}
\end{eqnarray}
which are spread evenly over the range $\theta_{0}\leq \theta_{\mu}\leq\theta_{0}+ 2\pi$, 
where $\theta_{0}$ is the initial (or reference) phase.

\section{Probability distributions}

Now given the form (\ref{correlatedpair}) and the definition of 
discrete phase states in section (\ref{phasesection}) we now proceed to calculate 
the joint probability of finding our correlated photon number system with phase 
$\theta_{\mu_{1}}$ for in the first subsystem/mode and $\theta_{\mu_{2}}$ for the 
second subsystem/mode. We specify that there is an initial phase angle $\theta_{0,i}$ 
for each of the subsystems $i$ (i=1,2). The joint probability 
$P_{\mu_{1},\mu_{2}}\left(\theta_{0,1},\theta_{0,2}\right)$ is given by
\begin{eqnarray}\label{generalsprob}
P_{\mu_{1},\mu_{2}}\left(\theta_{0,1},\theta_{0,2}\right)&=&
\sum_{n=0}^{s}  {{\left| c_{n}\right|^{2}}\over{\left(s+1\right)^{2}}}+ 
\sum_{n>n'=0}^{s}{{2 c_{n} c_{n'}}\over{\left(s+1\right)^{2}}} \nonumber \\
&\times& \cos \left[\Delta n
\left({{ 2\pi \left[ \mu_{1}+\mu_{2}\right]}\over{s+1}}+\psi_{0}\right)\right]
\end{eqnarray}
Here $\psi_{0}$ is the sum of the two initial phase angle 
($\psi_{0}=\theta_{0,1}+\theta_{0,2}$) and $\Delta n=n-n'$. 
Here each $\mu_{i}$ can vary from $0$ to $s$ in integer steps. 

It is useful to calculate the marginal probability $P_{\mu_{i}}\left(\theta_{0,i}\right)$ 
(where $i$ equals either 1 or 2) of finding the correlated photon number system with 
phase $\theta_{\mu_{i}}$ for the ${\rm i^{th}}$ subsystem (i=1,2), while having no 
information about the remaining subsystem (that is the  remaining subsystem 
may be in any phase state). Here the marginal probability 
$P_{\mu_{i}}\left(\theta_{0,i}\right)$ is given by
\begin{eqnarray}
P_{\mu_{i}}\left(\theta_{0,i}\right)= {1\over{s+1}}\sum_{n=0}^{s}  \left| c_{n}\right|^{2}
\end{eqnarray}
It is interesting to note that $P_{\mu_{i}}\left(\theta_{0,i}\right)$ is actually independent of all 
angular dependence, both from $\mu_{i}$ and $\theta_{0,i}$, the initial or reference choice 
of phase for that subsystem. In fact the marginal probability is uniformly 
distributed over all the possible results. In the limit that $s$ becomes very large 
$P_{\mu_{i}}\left(\theta_{0,i}\right)\rightarrow 0$. 

\section{Binary Choice}

We generally require large $s$ to describe an 
arbitrary phase for a general system. In fact to specify the phase as 
precisely as possible we require $s\rightarrow \infty$. 
However in the case of the measurement schemes  required for testing the various 
quantum paradox such as the Bell inequalities, all that is required and actually necessary
is a binary result \cite{binary choice}. Hence it would be logical 
to have a discrete phase measurement with say $s=1$, that is only two 
phase states. Such a scheme has been analyzed by Munro and 
Milburn\cite{Munro and Milburn 98} for the GHZ state. 

It may not always be possible to have only two phase states. 
If more phase states are present (for example $s=3$), a 
binary result is still required for these particular quantum 
paradoxes (and especially the Bell inequalities we are interested in, 
although some of the inequalities such as the Information theoretic 
Bell inequality\cite{Braunstein and Caves 1988} 
or the Mermin higher spin inequality\cite{Mermin90} allow for other 
than a binary result). 
This could be achieved by dividing or binning the phase states into two discrete distinct 
sets. We could for instance label $P_{\uparrow,\uparrow}\left(\theta_{0,1},\theta_{0,2}\right)$ 
as the probability of finding both particles in a $\uparrow$ state 
(where the $\uparrow$ state is one of the two possible binary results, the other being $\downarrow$). 
$P_{\downarrow,\downarrow}\left(\theta_{0,1},\theta_{0,2}\right)$ would 
 correspond to the probability of finding both modes in the $\downarrow$ 
state. How exactly this binary choice is achieved will be discussed on 
a per case basis in the next few sections of this paper. We could for instance 
specify that an $\uparrow$ result corresponds results containing the phase 
results $\mu= 0 \ldots s/2$. Such a process however discards 
information about the system and hence we would expect a lessening of our potential 
Bell inequality violation.

\section{The Bell inequalities}

\begin{figure}[h]
\center{ \epsfig{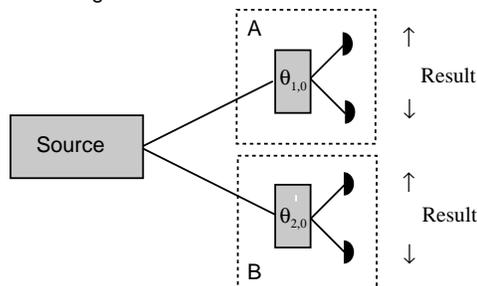}}
\caption{Schematic of a very generalized Bell experiment setup. After 
a source prepares two particles, these particles are directed out to 
the locations $A$ and $B$. At each location there is an analyzer with 
an adjustable parameters $\theta_{1,0},\theta_{2,0}$. The particles are then 
detected resulting in a binary result ``$\uparrow$'' or ``$\downarrow$'' individually.}
\label{fig1}
\end{figure}

A number of Bell inequalities exist, and the particular 
one to be considered in this article are the Clauser 
Horne\cite{Clauser and Horne 74} and the {\it spin}\cite{Bell65} Bell inequality.
A detailed derivation of the various inequalities will not be given, the 
reader is referred to references \cite{Bell65,Clauser and Horne 74}. 
In Fig (\ref{fig1}) we depict a very idealized setup for general Bell inequality experiment.

To formulate the Bell inequalities it is necessary to postulate the 
existence of a local hidden variable theory. We can write the probability 
$P_{\uparrow,\uparrow}\left(\theta_{0,1},\theta_{0,2}\right)$ 
for obtaining a result $\uparrow$ and $\uparrow$ respectively upon the 
simultaneous measurements the phase at $A$ and the phase at $B$ in terms 
of the hidden variables $\lambda$ as 
\begin{eqnarray}
P_{\uparrow,\uparrow}\left(\theta_{0,1},\theta_{0,2}\right)= \int \rho(\lambda) p_{\uparrow}^A(\theta_{1,0}, \lambda ) 
p_{\uparrow}^B(\theta_{2,0}, \lambda ) d\lambda  
\end{eqnarray}
The $\rho(\lambda)$ is the probability distribution for the hidden 
variable state denoted by  $\lambda$. $p_{\uparrow}^A(\theta_{0,2}, \lambda )$ 
is the probability of obtaining a result $\uparrow$ upon measurement at $A$ of 
the phase, given the hidden variable state {$\lambda$}. Similarly $p_{\uparrow}^B(\theta_{0,2}, \lambda )$ 
is the probability of obtaining a result $\uparrow$ upon measurement at $B$ of 
the phase, also given the hidden variable state  $\lambda$. 
Our locality  assumption is due to the fact that a  
measurement at $A$ cannot be influenced by the choice of parameter 
$\theta_{2,0}$ at the location $B$ (and vice versa).

A number of Bell inequalities exist and in this article we will 
consider only two cases. The first is the strong Clauser-Horne Bell inequality 
that can then be written in the form of 
\begin{eqnarray}\label{ch}
|{\bf B}_{\rm{CH}}|\leq 1
\end{eqnarray}
where
\begin{eqnarray}
B_{\rm{CH}}={{P_{\uparrow\uparrow}\left(\theta_{1,0},\theta_{2,0}\right) -P_{\uparrow\uparrow}
\left(\theta_{1,0},\theta_{2,0}'\right) 
+P_{\uparrow\uparrow}\left(\theta_{1,0}',\theta_{2,0}\right) +
P_{\uparrow\uparrow}\left(\theta_{1,0}',\theta_{2,0}'\right)}\over{ 
P_{\uparrow}\left(\theta_{1,0}'\right)+P_{1}\left(\theta_{2,0}\right)}}
\end{eqnarray}
Here $P_{\uparrow\uparrow}$ is the probability that a ``$\uparrow\uparrow$'' results occurs at each 
analyzer $A,B$ given $\theta_{1,0}$,$\theta_{2,0}$. Similarly $P_{\uparrow}$ is the probability 
that a ``$\uparrow$'' occurs at a detector while having no information 
about the second. For 
many of the actual experimental considerations an angle factorization 
occurs so that $P_{\uparrow\uparrow}\left(\theta_{1,0},\theta_{2,0}\right)$ depends only on 
$\theta_{1,0}+\theta_{2,0}$. Also $P_{\uparrow}\left(\theta_{i,0}\right)$ 
is independent of both $\theta_{1,0},\theta_{2,0}$. 
In this case $B_{\rm{CH}}$ can be simplified to
\begin{eqnarray}\label{chinequality}
	B_{\rm{CH}}={{2 P_{\uparrow\uparrow}\left( 
\psi\right)+ P_{\uparrow\uparrow}\left( -\psi\right)-P_{\uparrow\uparrow}\left(3 \psi\right)}\over{ 2 
P_{\uparrow}}}
\end{eqnarray}
where $\psi=\theta_{1,0}+\theta_{2,0}=-\theta_{1,0}'-\theta_{2,0}'=\theta_{1,0}+\theta_{2,0}'$ and 
$3\psi=\theta_{1,0}'+\theta_{2,0}$. In some cases it can be shown 
that 
$P_{\uparrow\uparrow}\left(\psi\right)=P_{\uparrow\uparrow}\left(-\psi\right)$ 
and hence this expression further simplifies to
\begin{eqnarray}
	B_{\rm{CH}}={{3 P_{\uparrow\uparrow}\left( \psi\right)-P_{\uparrow\uparrow}\left(3 \psi\right)}\over{ 2 
P_{\uparrow}}}
\end{eqnarray}
This is the commonly used form for ${\bf B}_{\rm{CH}}$. A violation of this strong Clauser 
Horne Bell inequality is possible if $|{\bf B}_{\rm{CH}}|> 1$.

The second form of the Bell inequality (sometimes referred to as the {\it 
spin} or original Bell inequality) is given by\cite{Bell65}
\begin{eqnarray}\label{spin}
|B_{\rm S}|\leq 2
\end{eqnarray}
where 
\begin{eqnarray}\label{spinx}
B_{\rm S}=|E\left(\theta_{1,0},\theta_{2,0}\right) &-&E\left(\theta_{1,0},\theta_{2,0}'\right) \nonumber \\
&+&E\left(\theta_{1,0}',\theta_{2,0}\right)+E\left(\theta_{1,0}',\theta_{2,0}'\right)|\leq 2
\end{eqnarray}
Here the correlation function $E\left(\theta_{1,0},\theta_{2,0}\right)$ is given by
\begin{eqnarray}\label{correlation}
E\left(\theta_{1,0},\theta_{2,0}\right)&=& 
P_{\uparrow\uparrow}\left(\theta_{1,0},\theta_{2,0}\right)+
P_{\downarrow\downarrow}\left(\theta_{1,0},\theta_{2,0}\right)\nonumber \\
&-&P_{\uparrow\downarrow}\left(\theta_{1,0},\theta_{2,0}\right)-
P_{\downarrow\uparrow}\left(\theta_{1,0},\theta_{2,0}\right)
\end{eqnarray}
As has been discussed above $P_{\uparrow\uparrow}$ is probability that 
a ``$\uparrow$'' results occurs at each analyzer $A,B$ given $\theta_{1,0},\theta_{2,0}$. 
$P_{\downarrow\downarrow}$ is probability that a ``$\downarrow$'' results occurs at each 
analyzer $A,B$, while $P_{\uparrow\downarrow}$ ($P_{\downarrow\uparrow}$) is probability that a 
``$\uparrow$'' (``$\downarrow$'') results occurs at the analyzer $A$ and 
a ``$\downarrow$'' (``$\uparrow$'') at $B$. With the angle factorization given above, the inequality (\ref{spin}) 
can be rewritten as
\begin{eqnarray}\label{spin1}
B_{\rm S}=2 E\left(\psi\right)+E\left(-\psi\right) -E\left(3 \psi\right) 
\end{eqnarray}
When $E\left(\psi\right)=E\left(-\psi\right)$ this expression further 
simplifies to
\begin{eqnarray}
B_{\rm S}=3 E\left(\psi\right) -E\left(3 \psi\right) 
\end{eqnarray}
A violation of this inequality is possible if $|{\bf B}_{\rm{S}}|>2$.

These are the two Bell inequalities that will be tested with our ideal 
correlated photon pairs and phase states.

\section{Binary Phase Measurements}

As has been mentioned above a logical choice for our discrete  
phase measurements would be to set $s=1$, that is a binary result.  For an initial state 
consisting only of a pair of correlated photon number states (given 
by (\ref{twostatesystem})) we have
\begin{eqnarray}
P_{\mu_{1}\mu_{2}}\left(\psi_{0}\right)={1\over 4}+{1\over{2}} c_{0} c_{1} \cos \left[ 
\left(\mu_{1}+\mu_{2}\right)\pi-\psi_{0}\right] 
\end{eqnarray}
We now specify that the result $\mu_{i}=0$ correspond to a $\uparrow$ measurement (one of our 
binary choices) while $\mu_{i}=1$ correspond to a $\downarrow$ measurement. 
Hence the joint probability of obtaining an $\uparrow\uparrow$ (or  
$\downarrow\downarrow$) result is
\begin{eqnarray}\label{p11}
P_{\uparrow\uparrow}\left(\psi_{0}\right)=P_{\downarrow\downarrow}\left(\psi_{0}\right)={1\over 4}+{1\over{2}} c_{0} c_{1} 
\cos \left[\psi_{0}\right] 
\end{eqnarray}
while the probability of obtaining an $\uparrow\downarrow$ 
($\downarrow\uparrow$) result is
\begin{eqnarray}
P_{\uparrow\downarrow}\left(\psi_{0}\right)=P_{\downarrow\uparrow}\left(\psi_{0}\right)={1\over 4}-{1\over{2}} c_{0} c_{1} 
\cos \left[\psi_{0}\right] 
\end{eqnarray}
The marginal probability for this case is simply given by $P_{\uparrow} = 
1/2$. It is then easy to calculate the expressions for $B_{\rm CH}$ 
and $B_{\rm S}$  and in fact we find
\begin{eqnarray}
{B_{\rm CH}}&=& {1\over 2}+{1\over 2} c_{0} c_{1} \left[ 3 
\cos \left(\psi_{0}\right) -\cos \left(3 \psi_{0}\right)\right] \\
{B_{\rm S}}&=& 2 c_{0} c_{1} \left[ 3 
\cos \left(\psi_{0}\right) -\cos \left(3 \psi_{0}\right)\right]
\end{eqnarray}
Setting $\psi_{0}=\pi/4$ to maximise the values of $B_{\rm CH}$ and 
$B_{\rm S}$ we have
\begin{eqnarray}\label{bellfinal1}
{B_{\rm CH}}&=& {1\over 2}+\sqrt{2} c_{0} c_{1} \\
{B_{\rm S}}&=& 4 \sqrt{2} c_{0} c_{1}
\end{eqnarray}
Our initial condition on $c_{0},c_{1}$ for normalization requires that 
$c_{0}^{2}+c_{1}^{2}=1$, and hence the maximum value of the product 
$c_{0}c_{1}$  is one half. Hence 
\begin{eqnarray}
{B_{\rm CH,max}}&=& {1\over 2}\left(1+\sqrt{2} \right) \\
{B_{\rm S,max}}&=& 2 \sqrt{2} 
\end{eqnarray}
which is leads to a violation of both inequalities. In fact it is the same size of 
violation as is obtained when single photon detector schemes are 
considered.

Above we have considered an initial state consisting only of a linear 
combination of two correlated photon states. An ideal parametric 
amplifier given by (\ref{correlatedpair}) with the $c_{n}$ coefficient 
specified by (\ref{parampcn}) actually has a infinity (or at least very large) 
number of correlated photon number pair states. It is nearly impossible via parametric amplification to achieve 
the simplified state considered previously with our choice of $c_{0}, 
c_{1}$. The paramp only produces the state given by (\ref{twostatesystem}) 
(with $c_{0}\gg c_{1}$) in the weakly pumped case. As the pump power 
increases higher order terms become significant. Generally in considering the 
typical photon detection Bell inequality schemes, 
the effect of the higher order terms of the form $|n\rangle|n\rangle$ (with $n>1$ has been to dramatically 
decrease (and eventually destroy) the violation\cite{munro and reid 
94,munro and reid 94a,munro and reid 94b}. Hence what is the effect of 
the higher order pairs in our discrete phase measurement scheme? 
How rapidly will the higher order terms destroy 
our violation.  In fact a quite surprising result occurs. It can be easily shown that the 
Bell inequality, with an arbitrary number of higher order pair correlated number state terms included, is  
\begin{eqnarray}
{B_{\rm CH}}&=& {1\over 2}+{1\over 2} {{c_{0} c_{1}}\over{c_{0}^{2}+c_{1}^{2}}} \left[ 3 
\cos \left(\psi_{0}\right) -\cos \left(3 \psi_{0}\right)\right] \\
{B_{\rm S}}&=&2 c_{0} c_{1} \left[ 3 
\cos \left(\psi_{0}\right) -\cos \left(3 \psi_{0}\right)\right]
\end{eqnarray}
Here we notice that the expression for ${B_{\rm S}}$ is unchanged for 
what we had indicated above and hence we will not analyse it further in 
this case. Focussing our attention on the expression for ${B_{\rm 
CH}}$ and optimising for angle we find
\begin{eqnarray}
{B_{\rm CH}}&=& {1\over 2}+\sqrt{2} {{c_{0} c_{1}}\over{c_{0}^{2}+c_{1}^{2}}}
\end{eqnarray}
The maximum is as previously provided $c_{0}=c_{1}$. We do not 
however has such a signient condition on $c_{0}$ which was needed 
when we have only two correlated pair states. Thus the addition of 
more states actually leads to a lessening of the condition for a violation. 
With the coefficient $c_{n}$ given for the parametric amplifier given 
by (\ref{parampcn}), large values of $\chi \epsilon \tau$ give 
$c_{0}\sim c_{1}$ and hence a maximal violation of the Bell inequality. 
The reason for this effect can be easily understood when it is noted that a binary phase measurement 
only involves the number states $|0\rangle$ and $|1\rangle$.

\section{Tests with larger number of phase states}

In our approach consider so far, we have examined a discrete (binary, $s=1$) 
phase measurement. Let us now consider the case where more phase 
states are present ($s>1$), an inefficient binary phase measurement. 
We have termed this an inefficient binary phase measurement because 
to formulate the Bell inequality it is generally necessary to bin the phase 
results into a binary category. Such a process discards information 
and hence we would not expect it to be as ideal as the binary phase measurement.

There are a number of ways that we can divide or segment our phase space 
measurements into two categories. Probably the 
simplest is to divide the phase space into two equal parts.  Our phase 
label $\mu$ has values from $0$ to $s$. We could define the 
$\uparrow$ bin to be contain phase results from $\mu=0$ to ${\rm 
Int}[s/2]$\footnote{By ${\rm Int}[s/2]$ we mean the integer value of 
s/2, eg. ${\rm Int}[3/2]=1$}
for  s odd ($s/2-1$ for s even)  and the $\downarrow$ 
bin to contain  phase results from ${\rm Int}[s/2]+1$ 
for s odd ( $s/2+1$ for s even) to $s$. For example if $s=3$ we say 
the $\uparrow$ bin corresponds to the sum of the $\mu=0,1$ phase 
results while the $\downarrow$ bin corresponds to the sum of 
$\mu=2,3$ results. If $s$ is even, then we have an odd number of phase states and we must decide 
which bin to put the extra phase result into. In the cases we are 
considering this extra phase state for $s$ even is assigned to the 
$\downarrow$ bin. For example with $s=2$, we have the following 
values possible values for $\mu$ (0,1,2). We say that the $\uparrow$ 
bin corresponds to the sum of the phase results from $\mu=0$, while the 
$\downarrow$ bin contains phase results for $\mu=1,2$. We note 
here that the extra phase state $\mu=1$ has been put in the $\downarrow$ bin. 
This choice is arbitrary. 

Another logical choice of binning 
could be to put only a single phase state into the $\uparrow$ bin, 
say $\mu=0$ while assigning all other phase results to the 
$\downarrow$ bin. For example if we consider $s=3$, we say the $\uparrow$ state corresponds 
to the $\mu=0$ phase results while the $\downarrow$ state corresponds 
to the sum of $\mu=1,2,3$ results. There are also a number of other 
choices for how this binning process may be achieved but we now focus our attention 
initially on the equal binning scheme.

\subsection{The equal binning scheme}

Let us consider the equal binning schemes with $s$ odd. 
As we have discussed above, for the equal binning scheme we define the 
$\uparrow$ bin to be contain phase results from $\mu=0$ to ${\rm Int}[s/2]$ and 
the $\downarrow$ bin to contain  phase results from ${\rm Int}[s/2]+1$ to $s$. For the Clauser 
Horne Bell inequality we need to determine two quantities, the 
probability $P_{\uparrow\uparrow}$ of obtaining an $\uparrow\uparrow$ 
result and $P_{\uparrow}$, the probability of knowing 
that an $\uparrow$ result occurred for one of the particles while 
having zero information about the seconds result. The 
probability $P_{\uparrow\uparrow}$ for obtaining an $\uparrow\uparrow$ 
result is  simply
\begin{eqnarray}
P_{\uparrow\uparrow}(\psi_{0})&=&\sum_{\mu_{1},\mu_{2}= 0}^{{\rm Int}[s/2]} 
P_{\mu_{1}\mu_{2}}(\psi_{0})
\end{eqnarray}
where $P_{\mu_{1},\mu_{2}}(\psi_{0})$ is given by Eqn (\ref{generalsprob}).
Similarly the marginal probability $P_{\uparrow}$ is given by
\begin{eqnarray}
P_{\uparrow}=\sum_{\mu_{1}= 0}^{{\rm Int}[s/2]}\sum_{\mu_{2}= 0}^{s} 
P_{\mu_{1},\mu_{2}}(\psi_{0})= {1\over 2} \sum_{n=0}^{s} \left| c_{n} \right|^{2}
\end{eqnarray}
For this particular case we have specified that the first particle is 
in the $\uparrow$ bin (which is why the sum over $\mu_{1}$ ranges 
from $0$ to ${\rm Int}[s/2]$) while having zero information about which bin the 
second particle is in (this is why the sum over $\mu_{2}$ ranges from $0$ to $s$).
Substituting  these probability expressions into the expression for $B_{\rm CH}$ given by (\ref{chinequality}) we have
\begin{eqnarray}
{B_{\rm CH}}={1\over 2}&+&{1\over{s+1}} \sum_{\mu_{1},\mu_{2}= 0}^{{\rm Int}[s/2]} \sum_{n>n'=0}^{s} 
{{c_{n} c_{n'}}\over{\sqrt{\sum_{m=0}^{s} \left| c_{m} 
\right|^{2}}}}\nonumber \\
&\times& \left\{ 2 \cos \left[\Delta n 
\left({{2\pi (\mu_{1}+\mu_{2}) 
}\over{s+1}}-\psi_{0}\right)\right]\right. \nonumber \\
&+&\cos \left[\Delta n \left({{2\pi (\mu_{1}+\mu_{2}) }\over{s+1}}
+\psi_{0}\right)\right] \nonumber \\
&-&\left.\cos \left[\Delta n \left({{2\pi (\mu_{1}+\mu_{2}) }\over{s+1}}
-3 \psi_{0}\right)\right]\right\}
\end{eqnarray}

For an equal superposition of correlated number states, that is 
$c_{n}$ is given by 
\begin{eqnarray}\label{cnequalchoice}
c_{n}=1/\sqrt{s+1},
\end{eqnarray}
we are now able to calculate $B_{\rm CH}$. The angles $\psi_{0}$ is 
chosen so as to optimise $B_{\rm CH}$. We need however to be careful 
with what values $\psi_{0}$ can range over. In fact $\psi_{0}$ is 
restricted to the range $[0,2 \pi/3]$ as our expression for $B_{\rm CH}$ 
also involves $3 \psi_{0}$. The angle must be chosen uniquely.

It is found for the equal binning case that only $s=1$ violates the 
inequality which is the result we obtained previously. No violation 
is possible for higher $s$. Hence we consider an alternate binning 
scheme in the next subsection.

\subsection{A single phase result in the $\uparrow$ bin}

An alternate choice for dividing the discrete phase space in two would 
be to choose a single phase state result to be in our $\uparrow$ bin. For 
simplicity we choose the $\mu=0$ phase state to be in our  $\uparrow$ 
bin. In this case  it is easy to show
\begin{eqnarray}
P_{\uparrow,\uparrow }\left(\psi_{0}\right)&=&\sum_{n=0}^{s}  {{\left| c_{n}\right|^{2}}\over{\left(s+1\right)^{2}}}
+{{2}\over{\left(s+1\right)^{2}}}\sum_{n>n'=0}^{s} c_{n} c_{n'}  \cos \left[\Delta n\psi_{0} \right]
\end{eqnarray}
and 
\begin{eqnarray}\label{Pupone}
P_{\uparrow}={1\over{s+1}}\sum_{n=0}^{s}  \left| c_{n}\right|^{2}.
\end{eqnarray}
The expression for $B_{\rm CH}$ is then given by 
\begin{eqnarray}
B_{\rm CH}={1\over {s+1}}+{{2}\over{s+1}} \sum_{n>n'=0}^{s}
{{c_{n} c_{n'}}\over{\sqrt{\sum_{m=0}^{s} \left| c_{m} \right|^{2}}}}
\left\{3\cos \left[ \Delta n \psi_{0} \right]\right. -\left.\cos \left[3 \Delta n  \psi_{0} \right] \right\}
\end{eqnarray}

\begin{figure}
\epsfig{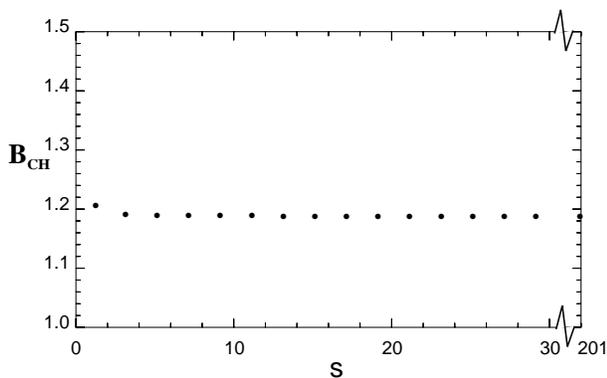}
\narrowtext
\caption{Plot of the ${B_{\rm CH}}$ versus s. A violation of the Bell 
inequality is possible for $|{B_{\rm CH}}|>1$.}
\label{fig2}
\end{figure}

Results for various odd $s$ are plotted in Fig  (\ref{fig2}). A violation of the Bell 
inequality is possible for $|{B_{\rm CH}}|>1$. We notice that as $s$ 
increases the violation does decrease but seems to be preserved for 
quite large $s$ (at least to $s=201$).

For the results displayed on Fig (\ref{fig2}) the $c_{n}$'s are given 
by (\ref{cnequalchoice}). The angle $\psi_{0}$ bounded in the range 
$[0,2 \pi/3]$ was chosen to maximise the value of $B_{\rm CH}$. It is 
interesting to note in this specific binning case that as $s$ increases, the 
width of the $\uparrow$ binning window decreases and hence the 
probability of finding it in that bin decreases. This is seen in Eqn 
(\ref{Pupone}). Here we observe that the $P_{\uparrow}$, the 
probability of detecting one subsystem in the $\uparrow$ bin while 
having zero information about the second, decreases as 
$P_{\uparrow}={1\over{s+1}}$. For large $s$ this probability becomes 
very small.

\begin{figure}
\epsfig{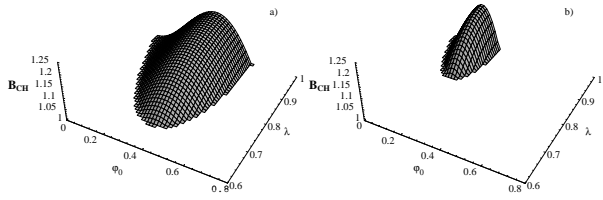}
\caption{Plot of the ${B_{\rm CH}}$ versus $\psi$ and $\lambda$ for 
$s=3$ (Figure a) and $s=7$ (Figure b).  A violation of the Bell 
inequality is possible for $|{B_{\rm CH}}|>1$.}
\label{fig3}
\end{figure}

The state considered above basically being an equal weighed superposition 
of the correlated pairs from $|0\rangle|0\rangle\ldots |s\rangle|s\rangle$ is extremely ideal. 
Let us briefly consider the effect to the potential violation if the 
$c_{n}$ coefficients are those for the ideal parametric amplifier. 
The $c_{n}$ coefficients were specified by eqn (\ref{parampcn}). For 
simplicity in the figure below we define a parameter 
$\lambda= \cosh \left[\chi \epsilon \tau \right]$ that varies from $0$ 
to $1$. $\lambda=0$ corresponds to only the state $|0\rangle|0\rangle$ being 
present while $\lambda=1$ corresponds to an equally weighed 
superposition. In Fig (\ref{fig3}), we plot $B_{\rm CH}$ versus $\psi_{0}$ 
and $\lambda$ for two values of $s$, $s=3$ and $s=7$.

The violation of the Bell inequality is seen in the Fig (\ref{fig3}) above as an 
{\it Island}. As $s$ increases we notice that the size of this island 
goes smaller. For even larger $s$ a violation is possible but the 
overall size of the {\it Island} of violation decreases significantly. 
$\lambda\rightarrow 1$ is the optimal choice of this parameter to 
maximise the violation but the angular dependence of $\psi_{0}$ does 
change with $s$.

Other binning choices are also available. We will not discuss these 
but some do lead to violations of the Bell inequality as seen above.

The last issue to be discussed in this section involves entanglement. 
By binning of the data we gain some insight on entanglement in two subsystem 
where each subsystem now has a binary state.  What our results show is 
that for some choices of binning there is enough entanglement left in 
the total system to violate  the strong Bell inequality. A question 
we pose but leave unanswered is whether a binned system could violate a 
Bell inequality but the original system not violate the corresponding 
multistate test of quantum mechanics.

\section{The spin analogy}
In this paper we have considered correlated  photon number pairs and 
discrete phase measurements. It is also possible to consider 
correlated spin systems. For instance we could write a correlated $s=1$ system as
\begin{eqnarray}
|\Psi\rangle= {1\over \sqrt{3}} \left[|1\rangle\otimes |1\rangle +
|0\rangle\otimes |0\rangle+|-1\rangle\otimes |-1\rangle\right]
\end{eqnarray}
Using similar binning schemes of spin measurements, it is possible to 
observe similar results to the phase measurements. Care does need to 
be taken in binning. There are optimal choices as was seen above.

\section{Conclusion}
Historically the Bell inequalities have restricted themselves to two 
physical subsystem where each subsystem has a binary state. Recent 
work by a number of authors have considered test with two subsystems, 
but where each subsystem has a larger number of states, sometimes infinite. In some  
of this work quadrature phase measurements were used  and the 
continuous results binned. In this paper we have considered correlated photon number 
pairs of the form $\sum c_{n} |n\rangle |n\rangle$ with a specific type of 
ideal discrete phase measurement. First we showed how binary phase 
measurements could produce a maximal violation of the Bell inequality. 
Then we considered cases where more than two phase results were 
possible. This was the primarily  purpose of the paper.

We showed how by binning the phase measurement results 
into two categories $\uparrow$ and $\downarrow$, 
a test of the Clauser Horne Bell inequality is possible (similar 
results do occur for the original Bell inequality). As was expected, the 
potential violation does decrease as the as the number of possible 
results from the phase measurements increases (as $s$ increases). 
This is to be expected as information in the phase measurement results
must be discarded to achieve the binary nature required for the test. 
We examined two specific binning cases with $s$ phase states. The 
first was to split the $s$ phase results into two equal sets. In this 
case a violation of the Bell inequality was only possible for $s=1$. 
Higher $s$ did not violate the inequality. The second case considered 
was where we took a single phase result ($\mu=0$) into one bin, 
with all the remaining possible results in the other bin. Such a 
binning scheme achieved a good potential violation of the inequality 
for quite high $s$ for a very idealised correlated state (in fact 
${1\over {s+1}} \sum_{n=0}^{s} |n\rangle |n\rangle$). The violation however did 
decrease as $s$ increases. When more realistic but still idealised 
states were considered (those from parametric down conversion), the 
effect of binning could truly be seen. A large violation was 
possible, but the parameter region over which a violation occurred 
decreased rapidly as $s$ increased. However for moderate values of 
$s$, this region is still quite large.

Finally while our results here have been applied to correlated photon number 
pair systems, they can be equally applied to higher spin systems. 
This opens the possibility for new novel tests of quantum mechanics. 
Insight can also be potentially about entanglement in this binned 
systems. What our results show for different types of binning is 
whether the system is entangled enough to be about to violate a 
strong Bell inequality.

\acknowledgments{WJM would like to acknowledge the support of the Australian Research Council.}

\begin{references}
\bibitem{strong} It is well known that some weakly entangled systems 
do not violate a Bell inequality and yet can not be produced 
classically. However strongly entangled systems do generally violate 
the Bell inequalities and hence the Bell inequality is a method of 
characterising the entanglement in such cases.
\bibitem{binary choice} Not all the inequalities used to test quantum 
mechanics require a binary choice. For example the Information 
theoretic Bell inequality requires a binary choice but the dimension 
of the subsystems is arbitrary.
\end{references}
\end{document}